\documentclass[11pt,a4paper]{article}

\usepackage[T1]{fontenc}
\usepackage[utf8]{inputenc}
\usepackage{lmodern}
\usepackage{microtype}
\usepackage{amsmath,amssymb}
\usepackage{graphicx}
\usepackage{booktabs}
\usepackage{placeins}
\usepackage{xurl}
\usepackage{hyperref}
\usepackage{url}
\usepackage{listings}
\usepackage{upquote}
\usepackage{xcolor}
\usepackage{tikz}
\usetikzlibrary{arrows.meta,positioning,calc,shapes.geometric}

\tikzset{
  flowbox/.style={
    rectangle, rounded corners=2pt, draw=blue!45!black,
    fill=blue!6, align=center, inner sep=4pt,
    minimum width=2.55cm, minimum height=0.72cm, font=\small
  },
  narrowbox/.style={
    flowbox, minimum width=2.05cm, minimum height=0.62cm, font=\footnotesize
  },
  widebox/.style={
    flowbox, minimum width=2.15cm
  },
  flowarr/.style={
    -{Stealth[length=2.2mm]}, semithick, draw=black!70
  }
}

\usepackage[numbers,sort&compress]{natbib}
\hypersetup{
  colorlinks=true,
  linkcolor=blue,
  citecolor=blue,
  urlcolor=blue,
  pdftitle={OpenMarket: A Synchronized Polymarket--Binance Dataset for High-Frequency Prediction-Market Research},
  pdfauthor={Gregory Young}
}

\graphicspath{{assets/figures/}}

\title{%
  OpenMarket: A Synchronized Polymarket--Binance\\
  Dataset for High-Frequency Prediction-Market Research
}

\author{%
  Gregory Young\\
  Department of Computer Science\\
  University of Colorado Boulder\\
  Boulder, CO 80309\\
  \texttt{gryo8540@colorado.edu}
}

\date{\today}

\IfFileExists{assets/stats/characterization.tex}{
\newcommand{\OpenMarketReleaseTag}{v0.5.2}
\newcommand{\OpenMarketTotalRows}{727,098,247}
\newcommand{\OpenMarketUnifiedGiB}{8.69}
\newcommand{\OpenMarketSnapshots}{202}
\newcommand{\OpenMarketLagPairs}{2,936,031}
\newcommand{\OpenMarketLagMedian}{16}
\newcommand{\OpenMarketLagPFive}{-186}
\newcommand{\OpenMarketLagPNinetyFive}{316}
\newcommand{\OpenMarketUnifiedMetaScanSeconds}{0.060}
\newcommand{\OpenMarketLagPairsLoadSeconds}{0.013}
\newcommand{\OpenMarketCollectionStart}{2026-03-14}
\newcommand{\OpenMarketCollectionEnd}{2026-07-01}

\newcommand{\OpenMarketPolyObservedDays}{54}
\newcommand{\OpenMarketBinanceObservedDays}{57}
\newcommand{\OpenMarketDataFirstDay}{2026-02-12}
\newcommand{\OpenMarketDataLastDay}{2026-05-15}
\newcommand{\OpenMarketDataSpanDays}{93}
\newcommand{\OpenMarketHardware}{Apple M5 Max, 64~GB RAM}
\newcommand{\OpenMarketRustVersion}{rustc 1.94.0 (4a4ef493e 2026-03-02) (Homebrew)}

\newcommand{\OpenMarketBenchBacktest}{0.09}
\newcommand{\OpenMarketBenchBacktestMarkets}{89}

}{
  \newcommand{\OpenMarketReleaseTag}{v0.5.2}
  \newcommand{\OpenMarketTotalRows}{727,098,247}
  \newcommand{\OpenMarketUnifiedGiB}{8.69}
  \newcommand{\OpenMarketSnapshots}{202}
  \newcommand{\OpenMarketLagPairs}{2,936,031}
  \newcommand{\OpenMarketLagMedian}{16}
  \newcommand{\OpenMarketLagPFive}{-186}
  \newcommand{\OpenMarketLagPNinetyFive}{316}
  \newcommand{\OpenMarketCollectionStart}{2026-03-14}
  \newcommand{\OpenMarketCollectionEnd}{2026-07-01}
  
  \newcommand{\OpenMarketPolyObservedDays}{54}
  \newcommand{\OpenMarketBinanceObservedDays}{57}
  \newcommand{\OpenMarketDataFirstDay}{2026-02-12}
  \newcommand{\OpenMarketDataLastDay}{2026-05-15}
  \newcommand{\OpenMarketDataSpanDays}{93}
}

\IfFileExists{assets/stats/research_stats.tex}{
\newcommand{\OpenMarketAucDiff}{0.0014}
\newcommand{\OpenMarketAucDiffLow}{0.0005}
\newcommand{\OpenMarketAucDiffHigh}{0.0023}
\newcommand{\OpenMarketAucPvalue}{0.001}
\newcommand{\OpenMarketLogisticAuc}{0.8415}
\newcommand{\OpenMarketNaiveAuc}{0.8401}
\newcommand{\OpenMarketDriftAuc}{0.7725}
\newcommand{\OpenMarketImbalanceAuc}{0.5863}
\newcommand{\OpenMarketLogisticEce}{0.027}
\newcommand{\OpenMarketNaiveEce}{0.014}
\newcommand{\OpenMarketLogisticScoreUs}{0.091}
\newcommand{\OpenMarketBootstrapN}{1,000}
\newcommand{\OpenMarketBootstrapBlocks}{2,251}
\newcommand{\OpenMarketOosRows}{355,814}
\newcommand{\OpenMarketNaiveOosAuc}{0.8405}
\newcommand{\OpenMarketNaiveOosBrier}{0.163}
\newcommand{\OpenMarketNaiveOosEce}{0.014}
\newcommand{\OpenMarketNaiveOosLogLoss}{0.485}
\newcommand{\OpenMarketModelOosAuc}{0.8377}
\newcommand{\OpenMarketModelOosBrier}{0.165}
\newcommand{\OpenMarketModelOosEce}{0.026}
\newcommand{\OpenMarketModelOosLogLoss}{0.495}
\newcommand{\OpenMarketFlagTightPct}{67.6}
\newcommand{\OpenMarketFlagMediumPct}{27.0}
\newcommand{\OpenMarketFlagWidePct}{5.5}
\newcommand{\OpenMarketEnvBinanceMed}{99}
\newcommand{\OpenMarketEnvBinanceRange}{4}
\newcommand{\OpenMarketEnvPolyMed}{7}
\newcommand{\OpenMarketEnvPolyRange}{2}
\newcommand{\OpenMarketDriftBound}{6}
\newcommand{\OpenMarketJumpCount}{15,148}
\newcommand{\OpenMarketJumpMatched}{4,272}
\newcommand{\OpenMarketRespMedianIngest}{347}

\newcommand{\OpenMarketBrierTight}{0.162}
\newcommand{\OpenMarketBrierWide}{0.185}
\newcommand{\OpenMarketSpreadMedian}{0.01}
\newcommand{\OpenMarketSpreadPNinety}{0.01}
\newcommand{\OpenMarketSpreadPNinetyFive}{0.02}
\newcommand{\OpenMarketSpreadPNinetyNine}{0.03}
\newcommand{\OpenMarketSpreadOneTickPct}{91.9}
\newcommand{\OpenMarketSpreadTwoTickPct}{6.0}
\newcommand{\OpenMarketLagPriceCorr}{0.00}

}{
  \newcommand{\OpenMarketAucDiff}{0.0014}
  \newcommand{\OpenMarketAucDiffLow}{0.0013}
  \newcommand{\OpenMarketAucDiffHigh}{0.0015}
  \newcommand{\OpenMarketAucPvalue}{<0.001}
  \newcommand{\OpenMarketLogisticAuc}{0.8415}
  \newcommand{\OpenMarketNaiveAuc}{0.8401}
  \newcommand{\OpenMarketDriftAuc}{0.7725}
  \newcommand{\OpenMarketImbalanceAuc}{0.5863}
  \newcommand{\OpenMarketBrierTight}{0.161}
  \newcommand{\OpenMarketBrierWide}{0.184}
  \newcommand{\OpenMarketSpreadMedian}{0.01}
  \newcommand{\OpenMarketSpreadPNinety}{0.01}
  \newcommand{\OpenMarketSpreadPNinetyFive}{0.02}
  \newcommand{\OpenMarketSpreadPNinetyNine}{0.03}
  \newcommand{\OpenMarketSpreadOneTickPct}{91.9}
  \newcommand{\OpenMarketSpreadTwoTickPct}{6.0}
  \newcommand{\OpenMarketLagPriceCorr}{0.00}
  \newcommand{\OpenMarketLogisticEce}{0.027}
  \newcommand{\OpenMarketNaiveEce}{0.015}
  \newcommand{\OpenMarketLogisticScoreUs}{0.102}
    \newcommand{\OpenMarketBootstrapN}{1,000}
    \newcommand{\OpenMarketBootstrapBlocks}{2,251}
  \newcommand{\OpenMarketOosRows}{355,814}
  \newcommand{\OpenMarketNaiveOosAuc}{0.8405}
  \newcommand{\OpenMarketNaiveOosBrier}{0.163}
  \newcommand{\OpenMarketNaiveOosEce}{0.014}
  \newcommand{\OpenMarketNaiveOosLogLoss}{0.485}
  \newcommand{\OpenMarketModelOosAuc}{0.8377}
  \newcommand{\OpenMarketModelOosBrier}{0.165}
  \newcommand{\OpenMarketModelOosEce}{0.026}
  \newcommand{\OpenMarketModelOosLogLoss}{0.495}
  \newcommand{\OpenMarketFlagTightPct}{67.6}
  \newcommand{\OpenMarketFlagMediumPct}{27.0}
  \newcommand{\OpenMarketFlagWidePct}{5.5}
  \newcommand{\OpenMarketEnvBinanceMed}{99}
  \newcommand{\OpenMarketEnvBinanceRange}{4}
  \newcommand{\OpenMarketEnvPolyMed}{7}
  \newcommand{\OpenMarketEnvPolyRange}{2}
  \newcommand{\OpenMarketDriftBound}{6}
  \newcommand{\OpenMarketJumpCount}{15,148}
  \newcommand{\OpenMarketJumpMatched}{4,272}
  \newcommand{\OpenMarketRespMedianIngest}{347}

}

\begin{document}
\maketitle

\begin{abstract}
OpenMarket began as an attempt to trade Polymarket's BTC 15-minute binary
markets against Binance BTC/USDT order flow. The attempt did not produce a
tradable edge: out-of-sample, a walk-forward logistic model over 43
microstructure features does not beat, and slightly underperforms, the
probability already implied by Polymarket's own order book, and simulated
trading nets
$-0.116$ normalized payoff units per attempted trade under stated fee and
slippage assumptions. We release
the synchronized corpus and infrastructure that attempt produced---to our
knowledge, the first public millisecond-level Polymarket BTC / Binance
BTC-USDT paired corpus with explicit pairing metadata. The frozen archive
(tag \OpenMarketReleaseTag{}) contains
\OpenMarketTotalRows{} deduplicated rows across \OpenMarketSnapshots{}
archival snapshots, with event data on \OpenMarketPolyObservedDays{}
observed Polymarket days (\OpenMarketBinanceObservedDays{} Binance days)
between \OpenMarketDataFirstDay{} and \OpenMarketDataLastDay{}, including
\OpenMarketLagPairs{} explicit
lead--lag pairs, alongside a reproducible Rust pipeline for collection,
millisecond pairing, Parquet export, and walk-forward calibration. Initial
analyses establish Polymarket stylized facts (one-tick top-of-book spreads)
and characterize cross-venue timing: an apparent \OpenMarketLagMedian{}~ms
median lag on venue source clocks, with relative clock drift bounded to
${\leq}\OpenMarketDriftBound{}$~ms over the archive but a remaining
single-vantage constant-offset ambiguity of roughly
$\pm\OpenMarketEnvBinanceMed{}$~ms. A synchronization-free event study, measured
only on the collector clock, independently shows that Polymarket quotes
respond to large Binance moves after a median \OpenMarketRespMedianIngest{}~ms.
We position this work as a data-and-methods release whose central empirical
result is a null out-of-sample forecasting result.
\end{abstract}

\section{Introduction}
\label{sec:introduction}

OpenMarket began as a trading system, not a dataset. We built a Rust
pipeline to trade Polymarket's BTC 15-minute binary markets against Binance
BTC/USDT spot flow: WebSocket collectors for both venues, a millisecond
recorder, a walk-forward calibrated logistic scorer, and simulated and paper
execution. The result was negative. No strategy iteration survived fees and
spread; out-of-sample, the strongest published model \emph{does not beat, and
slightly underperforms} the probability already implied by Polymarket's own
order book (Section~\ref{sec:insights-forecast}); and simulated positive-EV
trading nets $-0.116$ normalized payoff units per attempted trade under stated
fee and slippage assumptions. We do not claim trading profitability---the
central empirical finding is the opposite, and that negative result is
published with the data that produced it.

This paper releases the resulting infrastructure and corpus, which outlived
the trading strategy. Empirical work now documents tick-level
Polymarket dynamics, decentralized prediction-market (DePM) design
trade-offs, and combinatorial arbitrage
\cite{dubach2026anatomy,saguillo2025arbitrage,rahman2025sok}. What remains
missing is a \textbf{public, cross-venue, millisecond-resolution corpus}
paired with Binance BTC/USDT and tooling that makes synchronization
quality, lead--lag, and calibration analysis reproducible at archival
scale. OpenMarket closes that gap. The contribution is threefold:
\begin{enumerate}
  \item \textbf{Corpus.} To our knowledge, the first public millisecond-level
        Polymarket BTC / Binance BTC-USDT paired corpus with explicit pairing
        metadata: \OpenMarketTotalRows{} deduplicated rows across
        \OpenMarketSnapshots{} archival snapshots, with event data on
        \OpenMarketPolyObservedDays{} observed days
        (\OpenMarketDataFirstDay{}--\OpenMarketDataLastDay{};
        Section~\ref{sec:exp-scale}), including
        \OpenMarketLagPairs{} explicit lead--lag pairs.
  \item \textbf{Methods.} Documented source-vs.-ingest synchronization
        with empirical clock-offset validation
        (Section~\ref{sec:sync-clock}), Parquet-native export,
        walk-forward calibration training, and validation harnesses that
        treat clock drift and pairing-window sensitivity as measurable
        objects rather than implementation details.
  \item \textbf{Empirical baselines.} Stylized facts and forecast
        benchmarks on the released corpus---top-of-book spreads, lead--lag
        distributions, and ablations testing whether multivariate models
        outperform naive order-book mid priors (out-of-sample, they do
        not).
\end{enumerate}

The initial empirical claims are intentionally narrow:
\begin{enumerate}
  \item BTC 15-minute top-of-book spreads are usually one tick wide.
  \item Source-clock lead--lag has a compact median but heavy tails.
  \item Polymarket quotes respond to large Binance moves after hundreds of
        milliseconds on the collector clock.
  \item Lead--lag timing changes little across cross-venue disagreement regimes.
  \item Multivariate features slightly underperform the Polymarket mid prior
        out-of-sample.
\end{enumerate}

The released artifacts deliberately include the full research trail: the
collectors and recorder, the synchronization layer, the feature and
training pipelines, the backtesting framework, the strategy-evolution archive,
the Hugging Face dataset and model releases, and a sanitized operational
appendix for audit context. OpenMarket is frozen as a public research archive
(source tag \OpenMarketReleaseTag{}); active collection has ended. The remainder of the
paper describes the system, the released artifacts, and initial microstructure
findings that prior single-venue or closed pipelines cannot support.

\section{Background}
\label{sec:background}

Prediction markets aggregate information through prices of contracts tied to
future outcomes~\cite{wolfers2004prediction,hanson2003combinatorial}. Polymarket
lists binary outcome markets where contracts settle to 1 if an event occurs and 0
otherwise. BTC 15-minute markets ask whether Bitcoin will be above or below a
reference price at the end of a short window. These markets combine elements of
options, sports-style binary markets, and crypto exchange microstructure.
Recent Polymarket microstructure and DePM surveys are summarized in
Section~\ref{sec:related-work}. Reproducible public infrastructure for
high-frequency cross-venue work remains scarce.

Binance BTC/USDT serves as the reference stream because it is liquid,
high frequency, and tied to the Polymarket BTC settlement window. Following
multi-venue price-discovery practice~\cite{hasbrouck1995one}, we align Binance
trades with Polymarket book updates, derive features, generate labels, and
evaluate models without leaking future information. Millisecond-resolution
storage and pairing follow standard high-frequency microstructure
practice~\cite{ohara2015high,cont2014price}.
\section{Related Work}
\label{sec:related-work}

Prediction markets aggregate dispersed information into tradable prices
\cite{wolfers2004prediction,hanson2003combinatorial,berg2008prediction}.
The broader prediction-market literature studies forecast accuracy, incentive
compatibility, and event-study use cases, while market-microstructure research
studies how order flow, liquidity, latency, and price impact determine
short-horizon prices \cite{madhavan2000survey,ohara2015high,cont2014price}.
OpenMarket sits at the intersection: it treats a prediction market as a
high-frequency limit-order-book venue whose prices are probabilities.

\textbf{Public high-frequency order-book datasets.}
LOBSTER~\cite{huang2011lobster} reconstructs NASDAQ limit-order books for
academic use, and FI-2010~\cite{ntakaris2018benchmark} is a widely used
mid-price forecasting benchmark. The limit-order-book survey literature
\cite{gould2013limit} and neural LOB forecasting work such as
DeepLOB~\cite{zhang2019deeplob} show why public, timestamped order-book
benchmarks matter: they enable method comparison under fixed reconstruction and
evaluation rules. These datasets and models cover traditional equity venues.
OpenMarket follows this lineage but targets a decentralized prediction-market
CLOB paired with a crypto spot reference stream, settlement labels, and explicit
cross-venue pairing metadata.

\textbf{Polymarket and decentralized prediction-market data.}
Dubach~\cite{dubach2026anatomy} analyzes tick-level Polymarket order-book
evidence and releases analysis code, but does not ship a synchronized Binance
reference stream or a Hugging Face archival corpus. Saguillo et
al.~\cite{saguillo2025arbitrage} document combinatorial arbitrage and market
replications using proprietary-scale scrapes; reproduction requires
re-implementing data collection. Rahman et al.~\cite{rahman2025sok} survey DePM
design trade-offs without a cross-venue BTC 15-minute benchmark.
Recent Polymarket dataset releases are complementary rather than substitutes.
PolyBench~\cite{cheng2026polybench} provides timestamp-locked Polymarket order
book and news snapshots for LLM forecasting and trading evaluation. Jia et
al.~\cite{jia2026forecasting} construct a full-lifecycle Polymarket dataset
suite spanning market metadata, fill-level trading records, and oracle
resolution events. Qin and Yang~\cite{qin2026polymarketv1} release a complete
on-chain first-generation Polymarket archive with ground-truth aggressor
direction. OpenMarket differs by focusing narrowly on millisecond-level BTC
15-minute CLOB dynamics paired with Binance BTC/USDT and by publishing explicit
\texttt{lag\_pairs\_ms} construction metadata.

\textbf{Crypto price discovery and cross-venue timing.}
Multi-venue studies measure which market contributes to price discovery
\cite{hasbrouck1995one}. In crypto, Brandvold et
al.~\cite{brandvold2015bitcoin} estimate information shares across Bitcoin
exchanges, and Makarov and Schoar~\cite{makarov2020trading} document
cross-exchange crypto arbitrage and segmentation at scale. Public Binance kline
and trade products support many such studies, but they do not include
Polymarket CLOB state, settlement labels, or prediction-market-specific pairing
metadata. OpenMarket therefore does not claim to improve crypto price-discovery
estimation in general; it supplies a reproducible cross-venue corpus for a
specific prediction-market mechanism.

\textbf{Clock synchronization and timestamp validity.}
High-frequency cross-venue inference depends on timestamp semantics. NTPv4
specifies Internet clock synchronization against UTC~\cite{mills2010ntp};
precision-time deployments use IEEE 1588/PTP for tighter synchronization
\cite{ieee1588}. Financial regulation also treats clock accuracy as a
surveillance requirement: MiFID II RTS 25 specifies maximum UTC divergence and
timestamp granularity for trading venues and participants
\cite{mifid2016rts25}. OpenMarket is not a regulated trading venue and uses a
single public collector, so it cannot identify absolute venue-clock offsets
without additional vantage points. Its contribution is to publish both
source and ingest timestamps and to quantify which timing claims survive that
limitation.

\textbf{Open infrastructure.}
Most open-source trading repositories mix notebooks, private snapshots, and
undocumented assumptions. OpenMarket differs by publishing (i) raw and deduped
Parquet splits, (ii) explicit \texttt{lag\_pairs\_ms} with quality flags, (iii)
Rust exporters/trainers with pinned commands, and (iv) empirical baselines on the
released corpus.

We claim infrastructure and reproducible characterization, not superior live
trading performance. Dubach provides a deeper single-venue microstructure
analysis, while Saguillo et al. study broader proprietary-scale arbitrage data
than this public BTC 15-minute release.

\section{System Architecture and Data Collection}
\label{sec:architecture}

\begin{figure}[t]
  \centering
  \begin{tikzpicture}[
      node distance=0.42cm,
      every node/.style={flowbox}
    ]
    \node (bws) {Binance WS};
    \node[below=of bws] (tick) {Tick Stream\\Collector};
    \node[below=of tick] (sync) {Timestamp\\Synchronizer};
    \node[below=of sync] (feat) {Feature Generator};
    \node[below=of feat] (ml) {ML / Signal Engine};
    \node[below=of ml] (bt) {Backtester};
    \node[below=of bt] (eval) {Evaluation};

    \node[right=2.4cm of sync, minimum width=2.2cm] (pws) {Polymarket WS};
    \node[below=of pws] (book) {Order Book\\Collector};

    \draw[flowarr] (bws) -- (tick);
    \draw[flowarr] (tick) -- (sync);
    \draw[flowarr] (sync) -- (feat);
    \draw[flowarr] (feat) -- (ml);
    \draw[flowarr] (ml) -- (bt);
    \draw[flowarr] (bt) -- (eval);
    \draw[flowarr] (pws) -- (book);
    \draw[flowarr] (book.west) -- (sync.east);
  \end{tikzpicture}
  \caption{OpenMarket pipeline. Binance and Polymarket WebSocket streams feed
    collectors; the recorder synchronizes events and exports aligned tables for
    feature generation, signals, and backtesting.}
  \label{fig:architecture}
\end{figure}
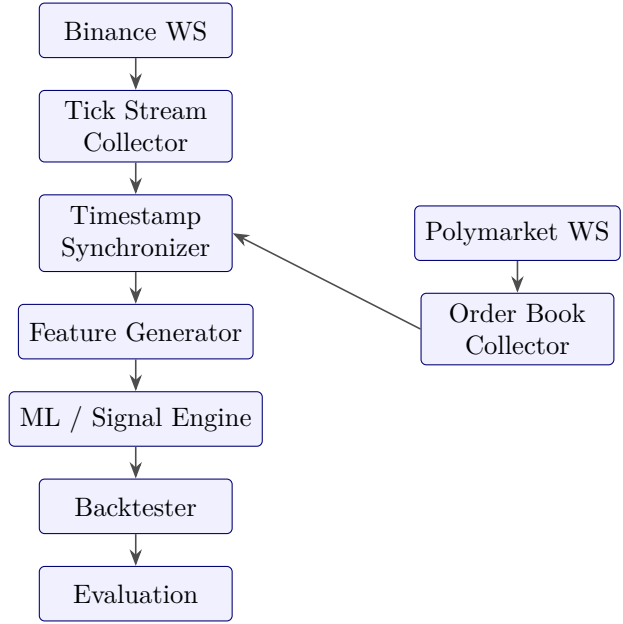

The codebase is a Rust workspace spanning exchange collectors, a multi-market
recorder with lag-pairing export, signal and execution engines, backtesting, and
Parquet-native ML crates (\texttt{step3-parquet-export},
\texttt{binary-outcome-trainer}); Figure~\ref{fig:architecture} shows how the
two WebSocket streams converge on the synchronizer and downstream stages.
Crate-level detail is maintained in the repository; the paper's benchmark
claims are regenerated from the frozen dataset and model artifacts, not from
live operational services.

\subsection{Binance}
\label{sec:data-binance}

The Binance collector records BTC/USDT trades and derives candle tables at
multiple time resolutions. The source schema includes trade ID, trade timestamp,
price, quantity, quote volume, maker/taker direction, and local receive time.
Millisecond-level tick snapshots preserve both source and ingest timestamps.

\subsection{Polymarket}
\label{sec:data-polymarket}

The Polymarket collector subscribes to BTC binary market order book updates,
trades, and last-trade-price events. The recorder maps token IDs to market slugs
and side labels using market metadata so that UP and DOWN books can be analyzed
consistently across rolling 15-minute markets.

\subsection{Storage}
\label{sec:data-storage}

The initial recorder stores data in SQLite for operational simplicity. The
public dataset release exports raw and processed tables to partitioned Parquet.
Key recorded tables include:
\begin{itemize}
  \item \texttt{binance\_trades}, \texttt{binance\_ticks\_ms}
  \item \texttt{polymarket\_ticks\_ms}, \texttt{market\_meta}
  \item \texttt{lag\_pairs\_ms}
  \item \texttt{binance\_candles\_1s}, \texttt{binance\_candles\_5s},
        \texttt{binance\_candles\_1m}, \texttt{binance\_candles\_5m},
        \texttt{binance\_candles\_15m}, \texttt{binance\_candles\_1h}
\end{itemize}


\section{Synchronization}
\label{sec:synchronization}

Synchronization is the central technical component of OpenMarket.
For each event the recorder stores both a source timestamp (exchange-emitted)
and an ingest timestamp (collector-observed):
\begin{align}
  \texttt{source\_ts\_ms} &= \text{timestamp from the exchange or event source}, \\
  \texttt{ingest\_ts\_ms}  &= \text{timestamp when the collector received the event}.
\end{align}
Following multi-venue price-discovery practice~\cite{hasbrouck1995one},
we pair Binance and Polymarket observations inside a bounded millisecond
window and compute lead--lag as
\begin{equation}
  \texttt{lead\_lag\_ms}
    = \texttt{polymarket\_source\_ts\_ms} - \texttt{binance\_source\_ts\_ms}.
  \label{eq:lead-lag}
\end{equation}
Positive values indicate that the Polymarket event timestamp follows the
Binance event timestamp; negative values indicate the opposite.
Note that \texttt{lead\_lag\_ms} is an \emph{event-timing} difference between
paired observations inside the window, not an econometric price-discovery
measure such as Hasbrouck's information share~\cite{hasbrouck1995one}; the
released pairs are the raw material from which such measures can be computed.
Each pair is stored with a quality flag (\texttt{tight}/\texttt{medium}/\texttt{wide}
pairing-window band), Binance price, Polymarket bid,
market slug, side label, and price delta in basis points.

\subsection{Pairing and Quality Metrics}

The canonical release constructs \texttt{lag\_pairs\_ms} with a
nearest-neighbor rule and alignment window $W=750$~ms. Polymarket ticks are
processed in recorder order in bounded batches. For each unpaired Polymarket
tick, the recorder queries Binance ticks whose source timestamps fall in
$[t_P-W,t_P+W]$, orders candidates by $|t_B-t_P|$, and stores the closest
candidate. Ties are broken by deterministic secondary ordering on
\texttt{binance\_source\_ts\_ms}, then the Binance tick row id; no random tie
breaking is used. The emitted lag is then $t_P-t_B$ as in
Equation~\eqref{eq:lead-lag}.

The pairing is one-to-one on the Polymarket side and many-to-one on the Binance
side: each Polymarket tick is marked paired after a match and is not reused, but
a high-rate Binance tick can be the nearest neighbor for multiple Polymarket
updates. This is deliberate because \texttt{lag\_pairs\_ms} is an event-alignment
table, not an order-book state reconstruction. The pairing code assigns
\texttt{quality\_flag} after matching from the realized absolute lag:
\texttt{tight} for $|\texttt{lead\_lag\_ms}|\leq100$~ms, \texttt{medium} for
$\leq300$~ms, and \texttt{wide} otherwise (within the 750~ms search window).
The public \texttt{v0.4.3-unified} split is produced later by merging overlapping
snapshot exports and deduplicating table partitions with table-specific keys;
for \texttt{lag\_pairs\_ms}, the dedupe key is
(\texttt{paired\_at\_ms}, \texttt{market\_slug}, \texttt{side\_label},
\texttt{binance\_source\_ts\_ms}, \texttt{polymarket\_source\_ts\_ms},
\texttt{polymarket\_bid}). Quality flags are therefore computed before unified
deduplication and preserved as row attributes.

OpenMarket treats the following as first-class dataset quality metrics,
consistent with microstructure guidance for decentralized prediction
markets~\cite{dubach2026anatomy,rahman2025sok}:
clock drift, dropped WebSocket messages, duplicate payloads, reconnect gaps,
stale order-book state, out-of-order events, and sensitivity to~$W$.

\subsection{Clock-Offset Validation}
\label{sec:sync-clock}

\begin{figure}[t]
  \centering
  \includegraphics[width=0.98\linewidth]{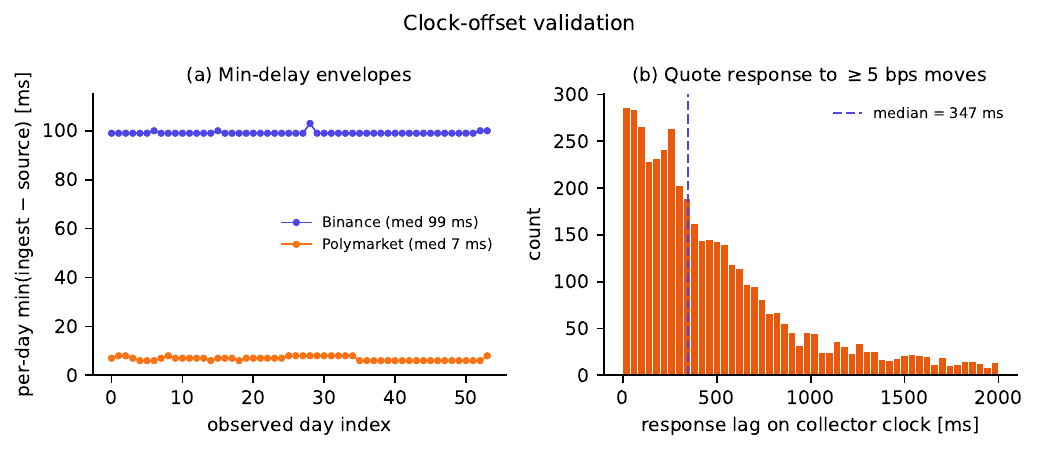}
  \caption{Clock-offset validation. (a) Per-day minimum transport-delay
    envelopes for both venues are flat across the archive, bounding relative
    clock drift to ${\leq}\OpenMarketDriftBound{}$~ms. (b) Distribution of
    Polymarket quote response lags after $\geq$5~bps Binance moves, measured
    on the collector's single clock (synchronization-free).}
  \label{fig:clock-validation}
\end{figure}

Equation~\eqref{eq:lead-lag} compares timestamps from two independent venue
clocks, so the lead--lag distribution is meaningful only up to their relative
offset and drift. With a single collection vantage point the \emph{constant}
component of the venue-clock offset cannot be separated from asymmetric
network latency; rather than assume it away, we measure what is identifiable
(Figure~\ref{fig:clock-validation}; regenerate with
\texttt{analyze\_research.py --refresh-clock}).

\textbf{Drift bound.} Every tick carries both venue and collector timestamps,
so per-venue transport delay $\texttt{ingest\_ts\_ms}-\texttt{source\_ts\_ms}$
is observable. Its per-day minimum envelope is
\OpenMarketEnvBinanceMed{}~ms for Binance and \OpenMarketEnvPolyMed{}~ms for
Polymarket, and each envelope varies by at most
\OpenMarketEnvBinanceRange{}~ms and \OpenMarketEnvPolyRange{}~ms respectively
across all observed days (Figure~\ref{fig:clock-validation}a). Any relative
drift between the venue clocks would move these envelopes; their stability
bounds cumulative relative clock drift to
${\leq}\OpenMarketDriftBound{}$~ms over the entire archive. The lead--lag
distribution is therefore not smeared by time-varying clock error; a constant
offset can only shift it uniformly. The same diagnostic does \emph{not}
identify the constant offset from a single collector host. The observed
minimum-delay envelopes imply a conservative single-vantage ambiguity of roughly
$\pm\OpenMarketEnvBinanceMed{}$~ms for source-clock lead--lag values, so the
\OpenMarketLagMedian{}~ms median should be read as an \emph{apparent}
source-clock median rather than an absolute cross-venue latency.

\textbf{Synchronization-free causal ordering.} To validate cross-venue
ordering without any venue-clock assumption, we detect
\OpenMarketJumpCount{} large Binance moves ($\geq$5~bps within 1~s) and match
\OpenMarketJumpMatched{} of them to the first directionally consistent
Polymarket UP-token best-bid change ($\geq$1 tick) within 2~s. Measured
entirely on the collector's single ingest clock---no cross-venue timestamp
comparison involved---the median quote response lag is
\OpenMarketRespMedianIngest{}~ms
(Figure~\ref{fig:clock-validation}b). This estimate uses only event ordering
on the collector clock, so the conclusion that Polymarket quotes
\emph{follow} large Binance moves does not depend on venue-clock offset.

We therefore report unconditional lead--lag medians as source-clock
quantities with drift bounded to ${\leq}\OpenMarketDriftBound{}$~ms and an
unresolved constant-offset ambiguity of roughly
$\pm\OpenMarketEnvBinanceMed{}$~ms; the causal response result is independent
of both.

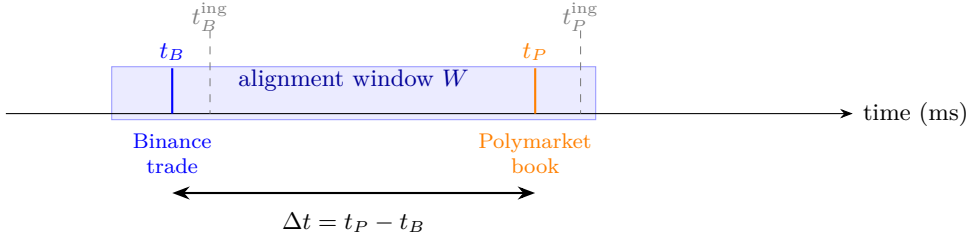
\begin{figure}[t]
  \centering
  \begin{tikzpicture}[
      >=Stealth,
      font=\footnotesize,
      ts/.style={thick},
      ingest/.style={dashed, gray},
      evlabel/.style={align=center, font=\scriptsize}
    ]
    \draw[fill=blue!8, draw=blue!40] (1.4,-0.08) rectangle (7.8,0.62);
    \node[blue!60!black] at (4.6,0.44) {alignment window $W$};

    \draw[->] (0,0) -- (11.2,0) node[right] {time (ms)};

    \draw[ts, blue] (2.2,0) -- (2.2,0.6);
    \node[blue] at (2.2,0.82) {$t_B$};
    \node[blue, evlabel] at (2.2,-0.52) {Binance\\trade};
    \draw[ingest] (2.7,1.1) -- (2.7,0);
    \node[gray] at (2.7,1.28) {$t_B^{\mathrm{ing}}$};

    \draw[ts, orange] (7.0,0) -- (7.0,0.6);
    \node[orange] at (7.0,0.82) {$t_P$};
    \node[orange, evlabel] at (7.0,-0.52) {Polymarket\\book};
    \draw[ingest] (7.6,1.1) -- (7.6,0);
    \node[gray] at (7.6,1.28) {$t_P^{\mathrm{ing}}$};

    \draw[<->, thick] (2.2,-1.05) -- (7.0,-1.05)
      node[midway, below=3pt] {$\Delta t=t_P-t_B$};
  \end{tikzpicture}
  \caption{Lead--lag pairing timeline. Source timestamps determine
    Equation~\eqref{eq:lead-lag}; ingest timestamps diagnose collector
    delay separately from cross-venue ordering.}
  \label{fig:lead-lag-timeline}
\end{figure}

\subsection{Lead--Lag Timeline (Figure~\ref{fig:lead-lag-timeline})}

Figure~\ref{fig:lead-lag-timeline} illustrates a single matched pair.
A Binance trade arrives at source time~$t_B$; a Polymarket book update
follows at~$t_P$. Ingest times ($t_B^{\mathrm{ing}}$, $t_P^{\mathrm{ing}}$)
are shown above the axis to separate network delay from exchange-reported
ordering. The shaded interval is the alignment window~$W$.

Empirical lead--lag distributions on the unified split ($n=\OpenMarketLagPairs{}$;
median \OpenMarketLagMedian{}~ms) are reported in
Section~\ref{sec:exp-sync} and Figure~\ref{fig:lead-lag-hist-unified}.

\FloatBarrier
\begin{samepage}
\subsection{Validation Checklist}

Researchers reproducing synchronization results should:
\begin{enumerate}
  \item verify monotonicity of \texttt{source\_ts\_ms} within each stream;
  \item count duplicate payloads and reconnect gaps;
  \item plot lead--lag histograms stratified by day and market;
  \item rebuild \texttt{lag\_pairs\_ms} from raw partitions and compare checksums.
\end{enumerate}
\end{samepage}

\section{Features and Machine Learning}
\label{sec:ml}

Feature families include order-book metrics (spread, best bid/ask, microprice,
imbalance, depth, book update velocity), price metrics (returns, realized
volatility, momentum, VWAP deviation, candle shape), technical indicators, and
custom short-horizon signals derived from aligned ticks. Features are paired
with settlement labels downstream. The repository also preserves exploratory
strategy and model prototypes, but the \textbf{published} binary-outcome model
uses a Rust pipeline on unified Parquet:

\begin{enumerate}
  \item \texttt{export\_step3\_from\_parquet} reads \texttt{v0.4.3-unified}
        tables and emits step3 calibration CSVs (43 features).
  \item \texttt{train\_binary\_outcome\_model} performs walk-forward logistic
        regression by market, applies Platt scaling~\cite{platt1999probabilistic},
        and writes JSON artifacts.
\end{enumerate}
The Polymarket mid prior used as the naive baseline is sampled at the same
feature-cutoff timestamp as the model inputs, so the comparison uses identical
information timing. Within each walk-forward window, both the logistic model
and Platt calibrator are fit only on pre-cutoff training/calibration rows;
scored OOS rows are not used for fitting or calibration.

On the publication workstation (Apple M5 Max), step3 export of 357{,}390 rows
across 2{,}251 markets completes in ${\sim}$63\,s; training (559 walk-forward
windows) completes in ${\sim}$67\,s. Step3 export writes 51\% of
\texttt{market\_meta} markets. The coverage audit over 4{,}450 markets reports
2{,}251 written, 1{,}241 with no Polymarket ticks in the export window, 922
with insufficient Binance trades for signal construction, 21 with missing
Binance partitions, 14 with no valid order-book snapshots after feature filters,
and one tied BTC close that is dropped because the binary label is undefined.

For public releases, model binaries are uploaded to Hugging Face Models rather
than committed to Git. The canonical \texttt{v0.2.1/} benchmark is the pooled
walk-forward OOS result across 559 windows: AUC \OpenMarketModelOosAuc{},
Brier \OpenMarketModelOosBrier{}, and ECE \OpenMarketModelOosEce{}. The same
rows give the naive mid prior AUC \OpenMarketNaiveOosAuc{}, Brier
\OpenMarketNaiveOosBrier{}, and ECE \OpenMarketNaiveOosEce{}, so the model
slightly underperforms the market prior. The frozen exported scorer evaluated on
the full 357{,}390-row step3 timeline is retained only as a diagnostic
reproducibility number because it includes each window's training rows
(Section~\ref{sec:insights-forecast}).
Simulated positive-EV trading under 1\% fees and 0.5\% slippage yields
$-0.116$ normalized payoff units per attempted trade. In the simulator, a
losing trade loses one staked unit plus fee, while a winning trade earns the
contract payoff net of entry price, fee, and slippage. An earlier
\texttt{v0.1/} pilot artifact remains for comparison.
Reliability and per-window metrics are regenerated by
\texttt{generate\_ml\_figures.py}; forecast ablations against naive baselines
are in Section~\ref{sec:insights-forecast}.

\section{Strategy, Backtesting, and Evaluation Framework}
\label{sec:strategy}

\subsection{Strategy Archive}

The repository preserves the trading-strategy lineage that motivated the data
release, but those iterations are archival context rather than paper claims.
They provide reproducibility hooks for simulated execution assumptions and
negative-result auditing. The paper uses them only to define the released
backtesting interface, fee/slippage assumptions, and model-evaluation metrics;
detailed strategy variants remain in the repository archive.

Position sizing and simulated execution assumptions are intentionally separated
from directional accuracy. Backtests report slippage, fees, assumed fill
prices, and market impact assumptions explicitly.

\subsection{Backtesting}
\label{sec:backtesting}

Backtesting processes market windows independently and evaluates entry signals
against settled outcomes under documented simulation assumptions (signal gate,
simulated fill, settlement, metric aggregation). Preferred
validation methods include walk-forward validation, rolling-window evaluation,
strict out-of-sample splits, sensitivity analysis over entry windows and
thresholds, and Monte Carlo resampling of fill and slippage assumptions. Random
row splits are discouraged because adjacent high-frequency observations are
highly autocorrelated.

\subsection{Evaluation Metrics}
\label{sec:evaluation}

Evaluation includes both predictive and simulated-economic metrics.
Economic metrics are counterfactual diagnostics under explicit fill, fee, and
slippage assumptions.

\paragraph{Predictive metrics.}
Accuracy, precision, recall, F1, ROC AUC, Brier score, and calibration
curves~\cite{brier1950verification}.

\paragraph{Simulated-economic metrics.}
Simulated win rate, expectancy, PnL, Sharpe and Sortino ratios (under stated
assumptions), maximum drawdown, turnover, average entry price, and documented
fill, fee, and slippage assumptions.

Calibration is especially important because binary market strategies are
sensitive to the difference between predicted probability and market-implied
price.

\section{Dataset}
\label{sec:dataset}

The public dataset is hosted on Hugging Face:
\url{https://huggingface.co/datasets/gregyoung14/openmarket-btc-polymarket}.

\subsection{Release Status}
\label{sec:dataset-status}

The frozen release has three dataset splits and two model artifacts: unified
\texttt{v0.4.3-unified} (the deduplicated analytic split), full
\texttt{v0.2-full} (per-snapshot archive), sample \texttt{v0.1-sample}
(quickstart), model \texttt{v0.2.1/} (canonical walk-forward logistic + Platt),
and model \texttt{v0.1/} (pilot comparison). The feature demo split is not a
full feature archive; full step3 features are reproducible from unified Parquet.

\begin{table}[t]
  \centering
  \caption{Version map for citing and reproducing OpenMarket artifacts.}
  \label{tab:artifact-versions}
  \footnotesize
  \setlength{\tabcolsep}{3pt}
  \begin{tabular}{@{}p{0.23\linewidth}p{0.19\linewidth}p{0.49\linewidth}@{}}
    \toprule
    Artifact & Version & Pins \\
    \midrule
    Source code & \OpenMarketReleaseTag{} & public launch commit, paper scripts, Rust pipeline \\
    Dataset unified split & \texttt{v0.4.3-unified} & deduplicated analytic Parquet timeline \\
    Dataset full split & \texttt{v0.2-full} & per-snapshot archival exports \\
    Dataset sample split & \texttt{v0.1-sample} & quickstart/demo subset \\
    Model release & \texttt{v0.2.1/} & walk-forward logistic + Platt artifact \\
    Model pilot & \texttt{v0.1/} & legacy comparison artifact \\
    \bottomrule
  \end{tabular}
\end{table}

Hugging Face's indexed/default dataset view may surface a larger aggregate row
count because it includes non-unified demo and full-archive artifacts; the
\OpenMarketTotalRows{} headline in this paper refers only to the deduplicated
\texttt{unified/} split.

The operator archive inventories \OpenMarketSnapshots{} CDN SQLite snapshots
published \OpenMarketCollectionStart{}--\OpenMarketCollectionEnd{}. The event
data inside those snapshots covers \OpenMarketPolyObservedDays{} observed
Polymarket days between \OpenMarketDataFirstDay{} and \OpenMarketDataLastDay{};
later snapshots are archival checkpoints of the same recorder state and add no
new event days (Section~\ref{sec:exp-scale}). Five formerly-partial
exports were recovered with \texttt{sqlite3 .recover} and re-exported before
the final unified rebuild; queue metadata reports 202~\texttt{published-clean},
0~\texttt{published-partial}, and 0~\texttt{corrupt}.

The unified split merges overlapping \texttt{full/} exports with
\path{scripts/datasets/merge_partitions.py}, removing ${\sim}$21\% duplicate
rows. This duplicate rate is expected: each \texttt{full/} snapshot is a
point-in-time export of recorder state, so adjacent snapshots intentionally
overlap in historical rows. New snapshots were published as archival checkpoints
of the current recorder database for reproducibility and recovery, rather than
as append-only deltas; this preserves independently valid per-snapshot exports
at the cost of duplicate historical rows in the raw union. The unified split
deduplicates those overlapping exports by table-specific keys and timestamps
rather than treating them as new events. All splits are validated with
\path{scripts/hf/validate_sample_split.py}. Empirical statistics are regenerated
from on-disk Parquet via \texttt{analyze\_unified.py} (Section~\ref{sec:reproducibility}).

Core table schemas are documented in the repository. Processed step3 features
for ML are exported from unified Parquet (\path{step3-parquet-export}) rather
than shipped as a full Hugging Face split.

\section{Experimental Characterization}
\label{sec:experiments}

We characterize the published \texttt{v0.4.3-unified} Hugging Face split with
the paper analysis scripts on local Parquet and model artifacts. Statistics below are computed from
on-disk Parquet metadata and a full scan of \texttt{lag\_pairs\_ms} (release tag
\OpenMarketReleaseTag{}). These results describe \textbf{what the archival corpus
contains}; microstructure findings and forecast benchmarks are in
Section~\ref{sec:insights}.

\subsection{Dataset Scale}
\label{sec:exp-scale}

\begin{figure}[t]
  \centering
  \includegraphics[width=0.92\linewidth]{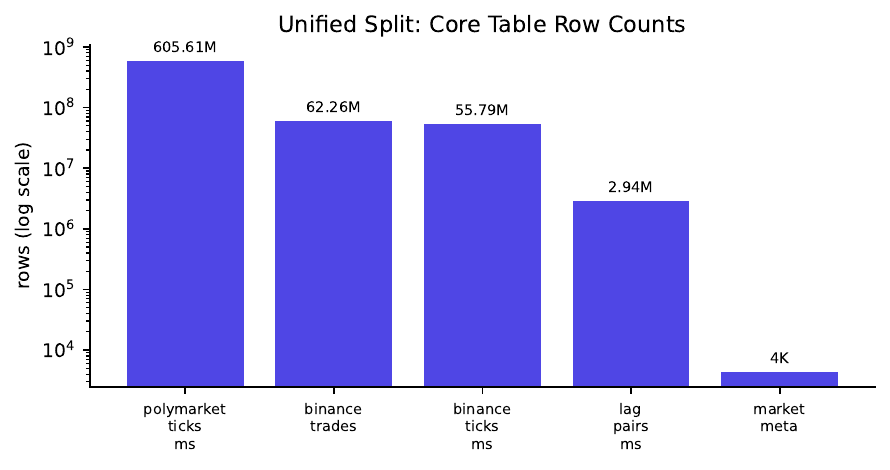}
  \caption{Unified split table row counts (log scale).}
  \label{fig:dataset-scale}
\end{figure}

The unified split contains \OpenMarketTotalRows{} rows across all exported tables
(Figure~\ref{fig:dataset-scale}),
occupying \OpenMarketUnifiedGiB{}~GiB on disk. The operator archive comprises
\OpenMarketSnapshots{} SQLite snapshots published from
\OpenMarketCollectionStart{} to \OpenMarketCollectionEnd{} (a 109-day
publication window). The event data inside those snapshots spans
\OpenMarketDataFirstDay{} to \OpenMarketDataLastDay{}
(\OpenMarketDataSpanDays{} calendar days), with event data on
\OpenMarketPolyObservedDays{} observed days for Polymarket and
\OpenMarketBinanceObservedDays{} for Binance; coverage is uneven, with
multi-day gaps, and no new event data was recorded after
\OpenMarketDataLastDay{} even though snapshot checkpoints
continued to be published until \OpenMarketCollectionEnd{}. After
deduplication across overlapping snapshot exports, the unified timeline removes
${\sim}$21\% duplicate rows relative to the raw \texttt{full/} union (916M
input rows $\rightarrow$ \OpenMarketTotalRows{} output rows). Most rows are
Polymarket ticks (605.6M), Binance trades (62.3M), Binance millisecond ticks
(55.8M), and explicit lag pairs (2.9M); candle tables add 498{,}525 rows and
have narrower coverage because only completed intervals are written.

The first five bars are the core event and metadata tables; derived candle
tables account for the remaining 498{,}525 rows in the headline total. Counts
and part numbers are tied to the frozen \texttt{v0.4.3-unified} manifest after
recovery and final deduplication; earlier draft tables generated from
pre-recovery subsets are not directly comparable.

Export quality: all \OpenMarketSnapshots{} snapshots are \texttt{published-clean}
after sqlite3 recovery of formerly-partial exports; per-snapshot export reports
document any table-level gaps.

\subsection{Synchronization Empirics}
\label{sec:exp-sync}

\begin{figure}[t]
  \centering
  \includegraphics[width=0.88\linewidth]{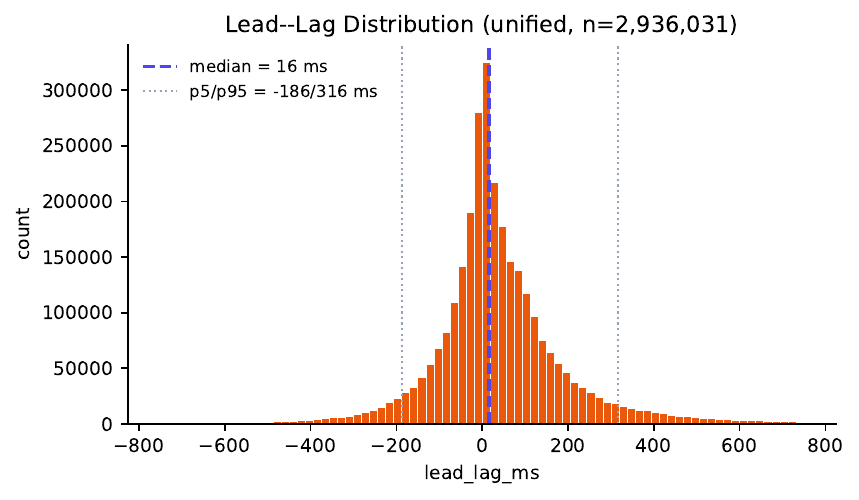}
  \caption{Empirical \texttt{lead\_lag\_ms} distribution from the unified
    \texttt{lag\_pairs\_ms} table ($n=\OpenMarketLagPairs{}$): the median is
    compact, but the distribution has heavy tails.}
  \label{fig:lead-lag-hist-unified}
\end{figure}

Lead--lag pairing produces \OpenMarketLagPairs{} matched event pairs in the
unified split. Figure~\ref{fig:lead-lag-hist-unified} shows the full distribution;
summary statistics are:

\begin{itemize}
  \item median lead--lag: \OpenMarketLagMedian{}~ms
  \item 5th / 95th percentiles: \OpenMarketLagPFive{} / \OpenMarketLagPNinetyFive{}~ms
\end{itemize}

Positive values indicate Polymarket source timestamps following Binance source
timestamps (Equation~\ref{eq:lead-lag}). On a 500k-pair sample, the recorder's
pairing-window quality flags split into \OpenMarketFlagTightPct{}\%
\texttt{tight} ($|$lag$| \le 100$~ms), \OpenMarketFlagMediumPct{}\%
\texttt{medium} ($\le 300$~ms), and \OpenMarketFlagWidePct{}\% \texttt{wide}
($> 300$~ms) pairs, so filtering to \texttt{tight} pairs retains roughly
two-thirds of the table. The long tails motivate treating
alignment-window sensitivity and clock drift as first-class quality metrics;
Section~\ref{sec:sync-clock} bounds relative clock drift empirically.

\subsection{Reproducibility Benchmarks}
\label{sec:exp-bench}

Environment-specific measurements on the publication workstation
(\OpenMarketHardware{}; \OpenMarketRustVersion{}) are recorded in
\texttt{assets/stats/benchmarks.json}. They include HF sample validation
(1.55\,s / 9{,}352 rows), unified Parquet metadata scan
(\OpenMarketUnifiedGiB{}~GiB in \OpenMarketUnifiedMetaScanSeconds{}\,s),
\texttt{lag\_pairs\_ms} load (\OpenMarketLagPairs{} rows in
\OpenMarketLagPairsLoadSeconds{}\,s), and the released staging backtest
(\OpenMarketBenchBacktest{}\,s across \OpenMarketBenchBacktestMarkets{}
markets). These are local I/O measurements, not cold-download benchmarks.

\section{Microstructure Findings}
\label{sec:insights}

This section reports \textbf{what we learn from the corpus}, not only what we
built. Regenerate via \texttt{analyze\_research.py} and
\texttt{analyze\_unified.py} (Section~\ref{sec:reproducibility}).

\subsection{Scale and Coverage}
\label{sec:insights-scale}

The unified archive contains \OpenMarketTotalRows{} rows with event data on
\OpenMarketPolyObservedDays{} observed Polymarket days between
\OpenMarketDataFirstDay{} and \OpenMarketDataLastDay{}
(Section~\ref{sec:exp-scale}).
Polymarket ticks dominate volume
(605.6M rows), enabling event studies at scales impractical with sample-only
public drops. After deduplicating overlapping snapshot exports, ${\sim}$21\% of
raw \texttt{full/} rows collapse---a methodological point for anyone merging
recorder archives without explicit keys.

\subsection{Lead--Lag and Cross-Venue Disagreement}
\label{sec:insights-leadlag}

\begin{figure}[t]
  \centering
  \includegraphics[width=0.98\linewidth]{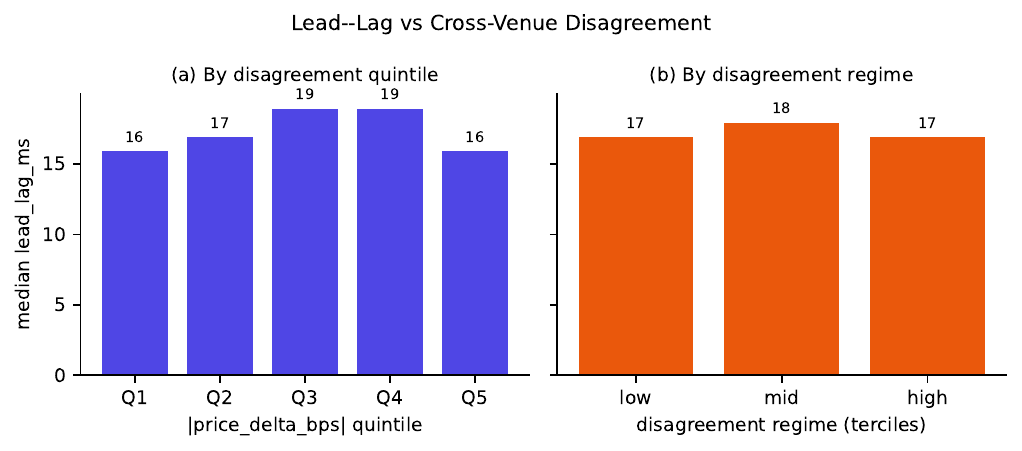}
  \caption{Median lead--lag by cross-venue disagreement. Across both
    $|$\texttt{price\_delta\_bps}$|$ quintiles and disagreement-regime
    terciles, median timing remains in a narrow 16--19~ms band
    (\texttt{lag\_pairs\_ms}, 500k subsample).}
  \label{fig:lead-lag-disagreement}
\end{figure}

Lead--lag pairs exhibit a compact apparent median (\OpenMarketLagMedian{}~ms
on venue source clocks) but heavy tails (5th/95th:
\OpenMarketLagPFive{} / \OpenMarketLagPNinetyFive{}~ms;
Figure~\ref{fig:lead-lag-hist-unified}). Section~\ref{sec:sync-clock} bounds
relative clock drift to ${\leq}\OpenMarketDriftBound{}$~ms, but leaves an
unresolved single-vantage constant-offset ambiguity of roughly
$\pm\OpenMarketEnvBinanceMed{}$~ms. The contemporaneous pairing median
measures continuous cross-venue flow; reactions to \emph{large} moves are
slower---Polymarket quotes take a median
${\sim}\OpenMarketRespMedianIngest{}$~ms to move one tick after a
$\geq$5~bps Binance move (Section~\ref{sec:sync-clock}). Stratifying by $|$\texttt{price\_delta\_bps}$|$
quintiles shows \textbf{stable} median lead--lag (16--19~ms) across disagreement
levels (Figure~\ref{fig:lead-lag-disagreement}a): cross-venue \emph{timing} does
not materially shift with contemporaneous basis-point divergence.
Disagreement-regime terciles show the same pattern
(Figure~\ref{fig:lead-lag-disagreement}b).

Daily event volumes generated by \texttt{analyze\_unified.py} show uneven
archive coverage without a simple ``high volume $\Rightarrow$ faster
cross-venue adjustment'' rule.
Lead--lag magnitude is uncorrelated with $|$\texttt{price\_delta\_bps}$|$
(Spearman $\rho \approx \OpenMarketLagPriceCorr{}$ on 500k pairs), so lag delay
alone does not predict contemporaneous mispricing magnitude.

\subsection{Polymarket Spread Stylized Facts}
\label{sec:insights-spread}

Top-of-book spreads on UP/DOWN tokens concentrate at one tick wide
(median \OpenMarketSpreadMedian{} probability points; 95th percentile
\OpenMarketSpreadPNinetyFive{}; Table~\ref{tab:spread-summary}). Tight spreads
imply that naive backtests using mid prices can overstate executable edge---consistent
with negative simulated PnL under fees and slippage (Section~\ref{sec:ml}).

\subsection{Forecast Benchmarks and Ablations}
\label{sec:insights-forecast}

\begin{figure}[t]
  \centering
  \includegraphics[width=0.92\linewidth]{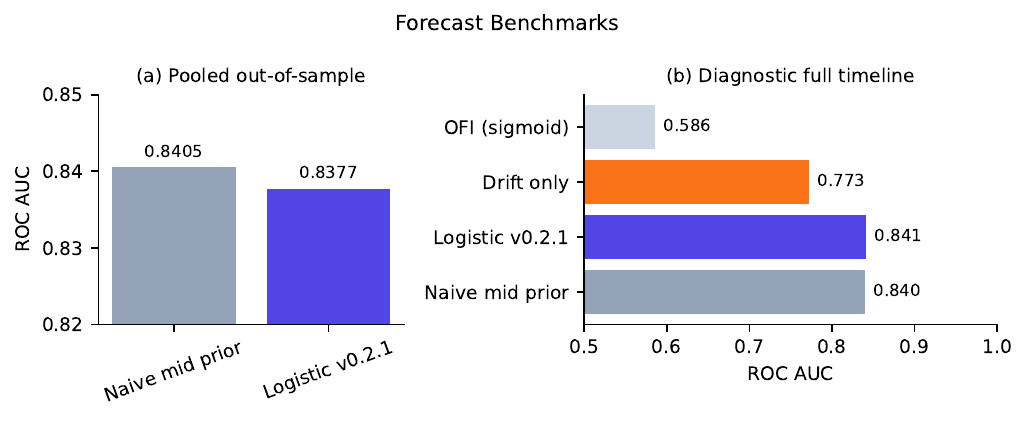}
  \caption{Forecast benchmark ROC AUC on step3 rows. Pooled out-of-sample
    evaluation is primary and shows the logistic model below the naive mid
    prior; full-timeline results are diagnostic only.}
  \label{fig:forecast-benchmarks}
\end{figure}

On 357{,}390 step3 rows exported from unified Parquet, we score the frozen
\texttt{v0.2.1} artifact and simple forecast baselines
(Table~\ref{tab:forecast-benchmarks}, Figure~\ref{fig:forecast-benchmarks}).
The naive mid prior is sampled at the same feature-cutoff timestamp as the
model inputs and evaluated on exactly the same scored rows.
All metrics are computed by \texttt{analyze\_research.py}; expected calibration
error (ECE) uses 10 equal-width probability bins~\cite{guo2017calibration}.

\begin{table}[t]
  \centering
  \caption{Forecast benchmarks on step3 rows. Pooled OOS rows are the primary
    comparison against the market prior. Diagnostic full-timeline rows include
    training rows from each walk-forward window and are reported only for
    reproducibility. Lower is better for Brier, ECE, and log loss.}
  \label{tab:forecast-benchmarks}
  \footnotesize
  \setlength{\tabcolsep}{3pt}
  \begin{tabular}{@{}llrrrrr@{}}
    \toprule
    Model & Split & AUC & Brier & ECE & Log loss & $\mu$s/row \\
    \midrule
    Naive mid prior & pooled OOS & \OpenMarketNaiveOosAuc{} & \OpenMarketNaiveOosBrier{} & \OpenMarketNaiveOosEce{} & \OpenMarketNaiveOosLogLoss{} & -- \\
    Logistic + Platt & pooled OOS & \OpenMarketModelOosAuc{} & \OpenMarketModelOosBrier{} & \OpenMarketModelOosEce{} & \OpenMarketModelOosLogLoss{} & -- \\
    Naive mid prior & diagnostic full & \OpenMarketNaiveAuc{} & 0.163 & \OpenMarketNaiveEce{} & 0.486 & -- \\
    Logistic + Platt & diagnostic full & \OpenMarketLogisticAuc{} & 0.163 & \OpenMarketLogisticEce{} & 0.489 & \OpenMarketLogisticScoreUs{} \\
    \texttt{drift\_prob\_up} only & diagnostic full & \OpenMarketDriftAuc{} & 0.218 & 0.145 & 0.792 & -- \\
    \texttt{imbalance\_60s} (sigmoid) & diagnostic full & \OpenMarketImbalanceAuc{} & 0.246 & 0.031 & 0.685 & -- \\
    \bottomrule
  \end{tabular}
\end{table}

\textbf{Effect size.} On the \OpenMarketOosRows{} rows the trainer scored
strictly out-of-sample, the naive prior attains AUC
\OpenMarketNaiveOosAuc{} versus the model's pooled OOS AUC of
\OpenMarketModelOosAuc{} (Brier \OpenMarketNaiveOosBrier{} vs.\
\OpenMarketModelOosBrier{}; ECE \OpenMarketNaiveOosEce{} vs.\
\OpenMarketModelOosEce{}): the logistic model does not beat, and slightly
underperforms, market-implied probability. For diagnostic context only, the full
timeline shows logistic \texttt{v0.2.1} above the naive mid prior by
$\Delta$AUC $= \OpenMarketAucDiff{}$ (paired market-block bootstrap over
\OpenMarketBootstrapBlocks{} market blocks, \OpenMarketBootstrapN{} resamples,
95\% CI $[\OpenMarketAucDiffLow{}, \OpenMarketAucDiffHigh{}]$; one-sided
$p=\OpenMarketAucPvalue{}$, $n=357{,}390$), but that comparison includes rows each
walk-forward window was trained on while the naive prior involves no training.
The in-sample ranking lift therefore does not survive walk-forward evaluation.

\textbf{Spread and reliability.} At one-tick spreads ($\leq 0.011$), Brier is
$\approx \OpenMarketBrierTight{}$; at wider quotes ($\geq 0.015$), Brier rises to
$\approx \OpenMarketBrierWide{}$. Realized-volatility terciles show a parallel
gradient (0.159 high-vol vs.\ 0.164 low-vol): forecasts are more reliable when
underlying BTC activity is elevated and degrade when Polymarket quotes widen.
Trade-direction features should also be treated as noisy public-feed
measurements: Dubach~\cite{dubach2026anatomy} reports only about 59\%
reliability for public-feed trade-direction inference, and our side labels and
\texttt{imbalance\_60s} inherit that limitation.

\textbf{Takeaway.} Multivariate features add at most a small diagnostic
full-timeline ranking lift over the Polymarket mid that does not survive
walk-forward evaluation; drift alone is weaker; raw order-flow imbalance fails
without context; and the
strongest published baseline remains economically dominated by fees, spread,
and calibration error. These are diagnostic baselines the infrastructure
enables at 357k-row scale in minutes on a laptop.

\begin{table}[t]
  \centering
  \caption{Polymarket top-of-book spread summary (200k tick subsample):
    most observations are one tick wide.}
  \label{tab:spread-summary}
  \footnotesize
  \setlength{\tabcolsep}{8pt}
  \begin{tabular}{@{}lr@{}}
    \toprule
    Statistic & Value \\
    \midrule
    Median & \OpenMarketSpreadMedian{} \\
    p90 & \OpenMarketSpreadPNinety{} \\
    p95 & \OpenMarketSpreadPNinetyFive{} \\
    p99 & \OpenMarketSpreadPNinetyNine{} \\
    Share at one tick ($\leq 0.011$) & \OpenMarketSpreadOneTickPct{}\% \\
    Share in two-tick band ($(0.011, 0.021]$) & \OpenMarketSpreadTwoTickPct{}\% \\
    \bottomrule
  \end{tabular}
\end{table}

\FloatBarrier
\section{Reproducibility}
\label{sec:reproducibility}

OpenMarket treats reproducibility as a first-class systems requirement. Any
benchmark, backtest, or paper result should report source commit, dataset
version, model version, config file, command, random seed, CPU, RAM, storage
type, OS, Rust version, and Python version. The repository ships three
reproduction paths: \texttt{cargo check --workspace} for build verification,
\texttt{scripts/hf/validate\_sample\_split.py} for the fast Hugging Face sample
split, and \texttt{datasets/download.py --split unified} plus the
\texttt{step3-parquet-export} and \texttt{binary-outcome-trainer} crates for
full model reproduction. Paper statistics and figures are regenerated by
\path{paper/scripts/compile.sh}; the clock-offset scan is cached in
\path{paper/assets/stats/clock_validation.json} and refreshed with
\texttt{analyze\_research.py --refresh-clock}. Legacy operator SQLite snapshots
remain available for migration only.

\section{Discussion}
\label{sec:discussion}

\subsection{What This Enables That Prior Work Does Not}

Dubach~\cite{dubach2026anatomy} establishes Polymarket microstructure stylized
facts on tick data; Saguillo et al.~\cite{saguillo2025arbitrage} quantify
arbitrage gaps. Neither provides a reproducible \textbf{Binance--Polymarket}
timeline with published pairing metadata and walk-forward calibration baselines.
OpenMarket lowers the cost of questions such as:
\begin{itemize}
  \item How stable is lead--lag across volatility/disagreement regimes?
  \item Does Polymarket mid probability subsume external drift signals?
  \item How sensitive are calibration metrics to fees, spreads, and pairing
        window~$W$?
\end{itemize}
Section~\ref{sec:insights} gives initial answers; the tooling is the enduring
contribution.

\subsection{Artifact Scope}

This work is best read as a \textbf{data-and-methods} release with empirical
baselines. Its primary claim is reproducible public infrastructure: a
synchronized corpus, pairing metadata, validation commands, and baseline
diagnostics that make future microstructure and forecasting comparisons easier
to audit. Its primary empirical result is negative: on this corpus, a
walk-forward multivariate model does not beat the market's own mid
out-of-sample, and simulated trading is cost-dominated.

\subsection{Domain Generalizability}

BTC 15-minute markets are a high-activity microcosm of Polymarket CLOB trading:
useful for synchronization methodology and short-horizon forecasting baselines,
but not representative of slower political or macro markets. Extensions to other
assets should preserve source/ingest timestamps and publish pairing metadata the
way this release does.

\subsection{Operational and Regulatory Context}

Polygon settlement latency, oracle definition risk, and U.S.\ regulatory
uncertainty affect economic interpretation but are outside our artifact. We
document clock-offset diagnostics and WebSocket gaps; users should not treat
top-of-book backtests as executable PnL without explicit queue and fee models.

\subsection{Data Availability and Ethics}
\label{sec:release}

The public release comprises the GitHub repository
(\url{https://github.com/gregyoung14/openmarket}, tag \OpenMarketReleaseTag{},
Apache~2.0), the Hugging Face dataset
\texttt{gregyoung14/openmarket-btc-polymarket} (\texttt{v0.4.3-unified},
\texttt{v0.2-full}, \texttt{v0.1-sample}) and models
\texttt{gregyoung14/openmarket-models} (\texttt{v0.2.1/}, \texttt{v0.1/}),
mdBook and Rust API documentation, a Jupyter quickstart
(\path{notebooks/quickstart.ipynb}), Docker reproducibility scaffolding, a
citation guide (\path{CITATION.md}), and the sanitized operational archive
(\path{research/operational/}).
The frozen source tag \OpenMarketReleaseTag{} is the immutable public-launch
source anchor; the GitHub release and tag page record the exact commit object.
The archive inventory contains \OpenMarketSnapshots{} snapshots
(46.21~GB compressed), and the generated characterization manifest is shipped
as \path{paper/assets/stats/characterization.json}. Until a DOI/Zenodo record
is assigned, derivative work should cite the Hugging Face dataset/model
revision hashes and the frozen source tag; a formal archival submission should
add that DOI alongside these version pins.
The corpus contains only public market data (order books, trades,
metadata) collected from open APIs---no user identities or private positions.
Dataset and model cards document licenses and version pins. Cite the dataset
version and source tag (\OpenMarketReleaseTag{}) in derivative work.

\subsection{Archive Status and Extensions}
\label{sec:future-work}

OpenMarket completed archival closeout on 2026-07-01. Active data collection
and model maintenance have ended; source tag \OpenMarketReleaseTag{} freezes
the public research record. Community extensions that the frozen artifacts
support include additional exchanges and assets, Polymarket non-crypto
markets, multi-vantage collection to resolve the constant clock-offset
component (Section~\ref{sec:sync-clock}), richer order-book reconstruction,
stronger model baselines (tree ensembles, neural sequence models), improved
execution simulation, and a public benchmark leaderboard---none of which
imply ongoing maintenance.

\section{Limitations}
\label{sec:limitations}

\textbf{Domain scope.} BTC 15-minute Polymarket binaries are a niche but liquid
regime: high tick rate, tight spreads, and volatile underlying spot. Stylized
facts and benchmarks here may not transfer to election markets, long-dated
events, or illiquid books without re-collection and re-calibration. In
particular, political and macro markets often have lower tick rates, wider and
more intermittent books, discrete news shocks, ambiguous or delayed settlement
mechanics, and no single liquid external reference stream analogous to
BTC/USDT. Those differences can break both the synchronization assumptions
(dense cross-venue observations inside millisecond windows) and the forecasting
features calibrated for short-horizon crypto price discovery.

\textbf{Data and methods.} WebSocket outages can create gaps; relative venue
clock drift is bounded empirically
(${\leq}\OpenMarketDriftBound{}$~ms, Section~\ref{sec:sync-clock}) but
source-clock lead--lag retains a roughly
$\pm\OpenMarketEnvBinanceMed{}$~ms constant-offset ambiguity from a single
collection vantage point; resolving that component would require
multi-vantage collection; top-of-book backtests may overstate execution
quality; mishandled
settlement labels can leak information; and step3 export covers 2{,}251 of
4{,}450 \texttt{market\_meta} markets for the reasons audited in
Section~\ref{sec:ml}. Full-archive \texttt{features/} Parquet is not on
Hugging Face (run \path{export_step3_from_parquet} locally).

\textbf{Model comparisons.} The repository contains legacy XGBoost, LightGBM,
stacking, and SHAP prototypes, but these were exploratory strategy artifacts,
not frozen benchmarks on the unified \texttt{v0.4.3} step3 export. The main
paper therefore reports only baselines regenerated from the archived release.
Direct tree-model, neural sequence, and market-by-market benchmark suites remain
future work.

\textbf{Live trading.} Simulated outcomes are sensitive to regime and fee
assumptions; live execution adds latency, queue position, partial fills, and
regulatory uncertainty not modeled in this artifact. The operational archive is
included only as sanitized audit context; it is not a recommendation to run live
trading infrastructure and should not be interpreted as a profitability claim.

\section{Conclusion}
\label{sec:conclusion}

We presented OpenMarket, an open-source pipeline and, to our knowledge, the
first public millisecond-level Polymarket BTC / Binance BTC-USDT paired corpus
with explicit pairing metadata. The release includes \OpenMarketTotalRows{}
deduplicated rows, \OpenMarketLagPairs{} lead--lag pairs, reproducible Rust
exporters and trainers, and Hugging Face artifacts
(\texttt{v0.4.3-unified}, \texttt{v0.2.1/} model).

Initial analyses support five narrow claims: BTC 15-minute books are usually
one tick wide at the top of book; source-clock lead--lag has a compact median
and heavy tails; large Binance moves precede Polymarket quote responses by
hundreds of milliseconds on the collector clock; lead--lag timing changes
little across disagreement regimes; and multivariate features slightly
underperform the naive mid prior out-of-sample. These are null and descriptive
results rather than trading-alpha claims. We invite researchers to use the corpus for
microstructure and forecasting studies, extend the tooling to new venues, and
cite the dataset version used in each experiment.

\section*{Funding and Competing Interests}
This project was produced as an independent open-source release. The author
declares no external funding and no competing interests.

\bibliography{bibliography}

\end{document}